\documentclass[12pt]{iopart}

\usepackage{iopams}
\usepackage{color}
\usepackage{soul}

\usepackage{graphicx} 
\newcommand {\be}{\begin{equation}}
\newcommand {\ee}{\end{equation}}
\newcommand {\ba}{\begin{eqnarray}}
\newcommand {\ea}{\end{eqnarray}}

\newcommand{\kl}{k_{_L}}
\newcommand{\kr}{k_{_R}}

\newcommand{\ko}{k_{_0}}
\newcommand{\Vl}{V_{_L}}
\newcommand{\Vr}{V_{_R}}

\newcommand{\nc}{n_{_c}}

\newcommand{\Tmpl}{T_{_L}}
\newcommand{\Tmpr}{T_{_R}}
\newcommand{\jif}{J_i^{(1)}}
\newcommand{\jis}{J_i^{(2)}}

\begin{document}


\title[]{Thermal rectification in segmented Frenkel-Kontorova lattices with asymmetric next-nearest-neighbor interactions}

\author{M.~Romero-Bastida and A.~Poceros~Varela}
\address{SEPI ESIME-Culhuac\'an, Instituto Polit\'ecnico Nacional, Av. Santa Ana No. 1000, San Francisco Culhuac\'an, Culhuac\'an CTM V, Coyoac\'an, CDMX 04440, Mexico}
\ead{mromerob@ipn.mx}

\date{\today}

\begin{abstract}
In this work we conduct an extensive study of the asymmetric heat flow, i.e. thermal rectification, present in the two-segment Frenkel Kontorova model with both nearest-neighbor (NN) and next-nearest-neighbor (NNN) interactions. We have considered systems with both high and low asymmetry and determined that, in the weak-coupling limit, thermal rectification is larger when NNN interactions are relevant. The behavior of the heat fluxes as a function of the coupling strength between the two segments is largely consistent with a well-defined rectification for larger system sizes. The local heat fluxes present a very different behavior for systems with high and low asymmetry. The results of this work may help in the design of molecular bridges, which have recently been shown to be able to function as thermal rectification devices.
\end{abstract}



\maketitle

\section{Introduction\label{sec:Intro}}

A thermal diode is an elementary device capable of sophisticated manipulation of heat flux. When reversing the temperature difference imposed on the two boundaries of such device, the forward and backward heat flows across the material can by asymmetrical, a phenomenon known as thermal rectification (TR). The first successful theoretical implementation~\cite{Terraneo02} was rapidly followed by proposals to use it as a foundation of a wide range of applications such as phononic transistors~\cite{Li06}, logical gates~\cite{Wang07}, and phononic memories~\cite{Wang08} among others. Furthermore, effective control of heat currents requires the development of nanoscale thermal devices that could efficiently harness waste thermal energy that hinders the performance of the smallest chips present in everyday electronic devices~\cite{Moore14}. Therefore, there has been a sustained effort to experimentally implement TR, such as by employing convection (thermosyphon)~\cite{Lee72}, by means of a semiconductor quantum dot~\cite{Scheibner08}, by phonon transport in a thin silicon membrane~\cite{Schmotz11}, or thermal expansion-contraction~\cite{Gaddam17}. However, the straightforward approach of connecting two materials with different temperature-dependent thermal conductivities remains the most promising alternative to construct a thermal rectifier that operates at room temperature, as exemplified by the implementation of a thermal rectifier by means of two coupled cobalt oxides~\cite{Kobayashi09}.

One of the first systems wherein a highly efficient asymmetric heat conduction was reported is a one-dimensional (1D) system composed of two coupled dissimilar Frenkel-Kontorova (FK) anharmonic lattices~\cite{Li04a}; the forward and backward heat fluxes differed by a factor of 100. In this work the explanation of this phenomenon was that the phonon bands of the different segments of the lattice change from overlap to separation when the heat reservoirs are reversed. However it was later determined that the rectification effect in this very same model is possible only in the weak interfacial coupling limit and is reversed when the properties of the interface and the system size change~\cite{Hu06}. Now, since the asymmetric heat conduction depends critically on the properties of the interface and system size, it was concluded that that it would be very difficult to fabricate a practical thermal diode from two coupled systems.

However, recent advances in nanomaterials and nanotechnology have, in fact, made fabrication of thermal diodes at micro- and nanoscales feasible, which affords new possibilities for high-efficiency thermal diodes~\cite{Maldovan13,Pop10}. Therefore, a fundamental understanding of the rectification mechanisms at small scales becomes critical. Several models have been proposed, including those based on asymmetric geometry of nanostructures~\cite{Han21,Ma18,Lee12}, asymmetric coupling with thermal contacts~\cite{Wu09}, and 1D oscillator chains with asymmetric mass distribution~\cite{Romero17,Yang07}. One of the most recent models where a significant TR has been observed correspond to molecular bridges (MBs) that covalently bond gold and carbon nanotubes~\cite{Dong19}. The performed analysis suggests that TR can be attributed to the mismatch of vibrational modes between neighboring sulfur and carbon atoms in the MBs. This result is particularly interesting, since it is essentially the same mechanism responsible of the rectification effect in the 1D asymmetric lattice with onsite potentials previously mentioned~\cite{Li04a}. Furthermore, the coupled 1D oscillator model could be considered as a very simplified model of the interface between the MB and the metal lead in these types of systems~\cite{Dinpajooh22,Wang22,Sharony20}. Therefore, it is important to study if any modification of the 1D model could improve its rectification efficiency since it could provide useful insights to improve the TR of the recently studied MBs.

In this paper we explore the effect on TR in the aforementioned 1D model of the inclusion of next-nearest-neighbor (NNN) interactions. With this structural modification it becomes feasible to study in a systematic way the effect of forces with significant magnitude beyond the nearest-neighbor (NN) range. The importance of this possibility is highlighted by the recently reported computation of the relative contributions of NN, NNN, and higher-order interactions to the overall energy flow in MBs composed of alkanedithiol molecules and gold substrates~\cite{Sharony20}. An additional motivation stems from the fact that the thermal transport properties of systems with such a type of interactions are significantly altered by their presence. For example, in a 1D oscillator lattice with randomly distributed NNN interactions ---being thus a model of a glass--- it has been determined that the observed energy localization increases the relaxation time of the system~\cite{RomeroArias08}. Furthermore, the presence of NNN interactions can change the size dependence of the thermal conductivity when the heat-carrying phonons are scattered by a quartic anharmonicity~\cite{Xiong12,Xiong14}. And finally, the asymmetric heat flow, i.e. TR, of some 1D oscillator systems has been shown to improve by the addition of those same interactions~\cite{Romero21}. Therefore, it is reasonable that there might be an appreciable influence of the NNN interactions on TR of the herein considered system.

The remainder of the paper is organized as follows: in Sec.~\ref{sec:Model} the model system and methodology are presented. Our results on the dependence of the TR on the structural parameters of the model in the presence of NNN interactions are reported in Sec.~\ref{sec:Res}. The discussion of the results, as well as our conclusions, are presented in Sec.~\ref{sec:Disc}.

\section{The Model\label{sec:Model}}

\begin{figure}\centering
\includegraphics[width=0.75\linewidth,angle=0.0]{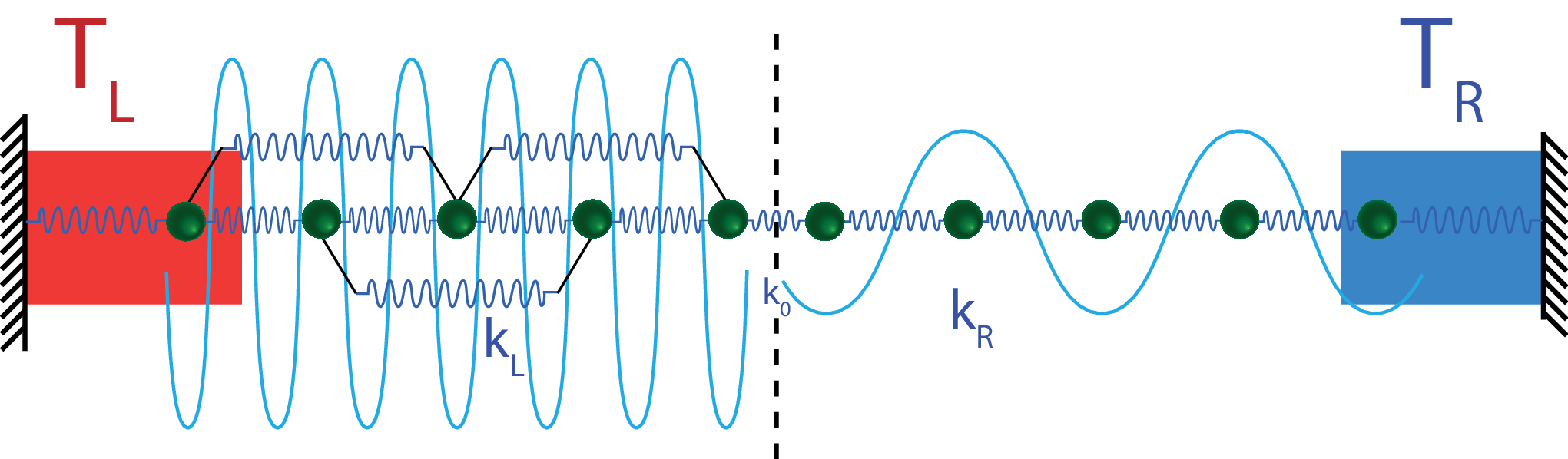}
\caption{Schematic representation of two dissimilar FK lattices, the left one with both NN and NNN interactions, connected by a harmonic spring. The whole system is attached at both ends to a thermal reservoir, each operating at different temperatures.}
\label{fig:1}
\end{figure}

The herein considered system consists of a size $N$ nonlinear oscillator lattice consisting of two segments ($L,R$) of size $\nc=N/2$ coupled together by a harmonic spring with constant strength $\ko$; a sketch of the model is presented in Fig.~\ref{fig:1}. The equations of motion (EOM) of any given oscillator within each segment can be written, in dimensionless variables, as $\dot q_i =p_i/m_i$ and 
\ba
\dot p_i& = &F^{(1)}_i + \gamma\sum_{j=1}^{\nc}F^{(2)}_i\delta_{ij} - {V_{i}\over2\pi}\sin(2\pi q_i/a) \cr
   & + & (\xi_{_1} - \alpha_{_1} p_i)\,\delta_{i1} + (\xi_{_1} - \alpha_{_N} p_i)\,\delta_{i{N}},
\ea
being $F^{(1)}_i=k_{i-1}(q_{i-1}-q_i)+k_i(q_i-q_{i+1})$ and $F^{(2)}_i=k_{i-2}(q_{i-2}-q_i)+k_i(q_{i+2}-q_i)$. Furthermore,
\ba
k_i&=&\sum_{j=1}^{\nc-1}\kl\delta_{ij} + \ko[\delta_{i \nc}+\delta_{i (\nc+1)}] +\!\!\!\! \sum_{j=\nc+2}^N\!\!\!\kr\delta_{ij}, \cr
V_i&=&\sum_{j=1}^{\nc}\Vl\delta_{ij} + \!\!\sum_{j=\nc+1}^N\!\!\Vr\delta_{ij},
\ea
where $k_{_{L,R}}$ and $V_{_{L,R}}$ are the harmonic spring constant and amplitude of the FK onsite potential in each segment, respectively. $\ko$ is the harmonic spring constant that connects both sides of the system. $\{m_i,q_i,p_i\}_{i=1}^{N}$ are the dimensionless mass, displacement, and momentum of the $i$th oscillator. The tunable parameter $\gamma$ specifies the relative strength of the NNN interaction $F^{(2)}_i$ compared to the NN one $F^{(1)}_i$ on the left side. Just as done in previous works~\cite{Li04a,Romero20} we set $\Vr=\lambda\Vl$ and $\kr=\lambda\kl$ in order to reduce the number of adjustable parameters. The above EOM corresponds to the commensurate case ---which is the only one considered in this study--- where the onsite potential assumes the same spatial periodicity as the lattice constant $a=1$. Henceforth we will consider a homogeneous system, i.e., $m_i=1\,\,\forall\,i$ and fixed boundary conditions ($q_{_0}=q_{_{{N}+1}}=0$). The stochastic force $\xi_{_{1,N}}$ is a Gaussian white noise with zero mean and correlation $\langle\xi_{_{1,N}}(t)\xi_{_{1,N}}(t^{\prime})\rangle=2\alpha_{_{1,N}}k_{_B}T_{_{1,N}}m_i(\delta_{1i}+\delta_{{N}i})\delta(t-t^{\prime})$, being $\alpha_{_{1,N}}$ (taken as $0.5$ in all computations hereafter reported) the coupling strength between the system and the left and right thermal reservoirs operating at temperatures $T_{_L}=0.105$ and $T_{_R}=0.035$, respectively; the system thus operates at a constant average temperature value of $T_{_0}\equiv(T_{_L}+T_{_R})/2=0.07$. This system could be considered as a very simplified model of a molecular junction. Starting from an initial configuration wherein the oscillators are in their equilibrium positions and the momenta are drawn from a Gaussian distribution compatible with the chosen average temperature $T_{_0}$ the complete set of EOM were integrated with a stochastic velocity-Verlet integrator with a timestep of $10^{-2}$ for a stationary time interval of $2\times10^7$ time units after a transient interval of $10^{8}$ time units.

Once the non-equilibrium stationary state is attained, the local heat flux is computed as
\ba
J_i& =\langle\dot q_{i} k_i(q_i-q_{i+1})\rangle + 2\gamma\sum_{j=1}^{\nc-2}\langle\dot q_{i} k_i(q_i-q_{i+2})\rangle\delta_{ij} \nonumber \\
   & =\jif + 2\gamma\sum_{j=1}^{\nc-2}\jis\delta_{ij}, \label{lhf}
\ea
with $k_i=\kl$ if $i\in[2,\nc-1]$, $k_i=\ko$ if $i=\nc$, and $k_i=\kr$ if $i\in[\nc+1,N]$. The local temperature is computed as $T_i=\langle p_i^2/m_i\rangle$. In both instances $\langle\cdots\rangle$ indicates time average over the entire time interval corresponding to the stationary state, in which the heat flux within each segment becomes independent of the site, and thus, in order to improve the statistical precision of our results, the mean heat flux $J$ is calculated as the algebraic average of ${J}_i$ over the number of unthermostatted oscillators with a precision of of $\mathcal{O}(10^{-6}-10^{-8})$. By $J_+$ we denote the heat flux when the high temperature reservoir is attached to the left end of the system and by $J_-$ the flux when that same reservoir is now connected to the opposite end of the lattice, i.e., the positions of the reservoirs are interchanged. With the quotient $r\equiv|J_+/J_-|$ we quantify the rectification efficiency of this device.

\section{Results\label{sec:Res}}

\subsection{Dependence on the harmonic coupling strength}

\begin{figure}\centering
\includegraphics[width=0.65\linewidth,angle=0.0]{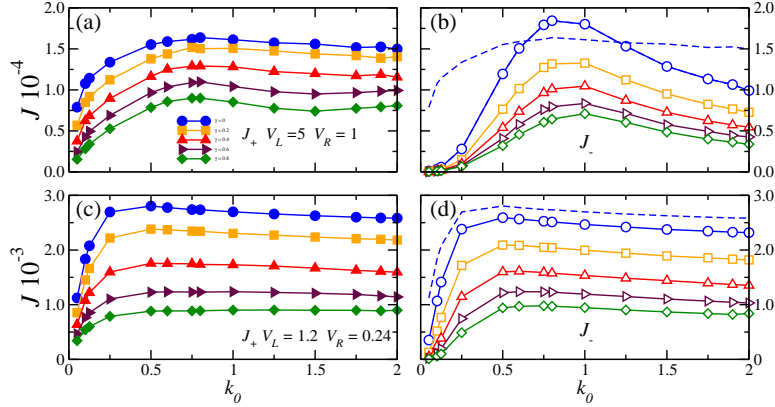}
\caption{(a,b) Heat flux vs the interfacial harmonic coupling $\ko$ for various values of relative strength of the NNN interactions: $\gamma=0$ (circles), 0.2 (squares), 0.4 (triangles up), 0.6 (triangles right), and 0.7 (diamonds), with $\Vl=5$; the same for panels (c,d), but with $\Vl=1.2$. Dashed lines are the corresponding $\gamma=0$ cases of panels (a,b). Solid and void symbols correspond to the forward and reverse-bias configurations, respectively. In all panels $N=100$, $\kl=1$ and $\lambda=0.2$. Error bars are smaller than symbol size. Continuous lines are a guide to the eye.}
\label{fig:2}
\end{figure}

Fig.~\ref{fig:2} presents the dependence of $J_{\pm}$ on $\ko$ for a fixed system size of $N=100$. In panels (a) and (b) the corresponding results of Ref.~\cite{Hu06} for $\Vl=5$ in the absence of NNN interactions are presented for comparison. It is clear that, even for low values of the relative strength of the NNN interactions, $J_+>J_-$ in the full range of $\ko$ values considered, contrary to the case with NN interactions only where there is a region between $0.6<\ko<1.2$ in which $J_->J_+$, i.e., the rectification effect is reversed. In Ref.~\cite{Hu06} it was argued that this phenomenology indicates that both sides of the system are no longer separable and that the explanation based in the phonon band overlap breaks down. However, our results seem to indicate that, in the presence of NNN interactions, both sides of the system remain separable and the aforementioned explanation remains valid. The magnitude of both heat fluxes tends to decrease in comparison to those for $\gamma=0$ and, for high $\ko$ values, $J_+$ is almost independent of the harmonic coupling value whereas $J_-$ still shows a significant dependence on $\ko$, at least for the low $\gamma$ values considered. For the larger $\gamma$ values there is a mild dependence of $J_+$ on $\ko$ and both fluxes tend to have a more similar behavior in the whole value range of $\ko$. It seems that the increased asymmetry afforded by the NNN interactions renders a more controllable dependence of the heat fluxes on the coupling strength value. Now, it is important to remark that two asymmetries are at play: that of the NNN interactions in the left side and the one in the magnitude of the onsite potential already considered. In order to determine if both are necessary to obtain the desired behavior in panels (c) and (d) we repeat the same computations, but now with a value of $\Vl=1.2$, which reduces the asymmetry on both sides of the system. The magnitude of the fluxes increases in an order of magnitude for all considered cases and, for the lowest $\gamma$ values considered in panels (c) and (d), $J_+>J_-$ in the entire $\ko$ value range, just as in the previous instance. However, it is also evident that the behavior of both fluxes is very similar; thus reduced rectification values are expected. For the larger $\gamma$ values considered the behavior $J_->J_+$ is obtained. Nevertheless we remark that this effect is greatly reduced in magnitude in comparison to the one corresponding to $\gamma=0$ and $\Vl=5$, first reported in Ref.~\cite{Hu06}. This reversal of the rectification effect thus seems a result of the high asymmetry afforded by $\Vl=5$ since it is only obtained, in the case of the $\Vl=1.2$ instance, for high $\gamma$ values, which again increase the asymmetry of the system.

\begin{figure}
\centerline{\includegraphics*[width=75mm]{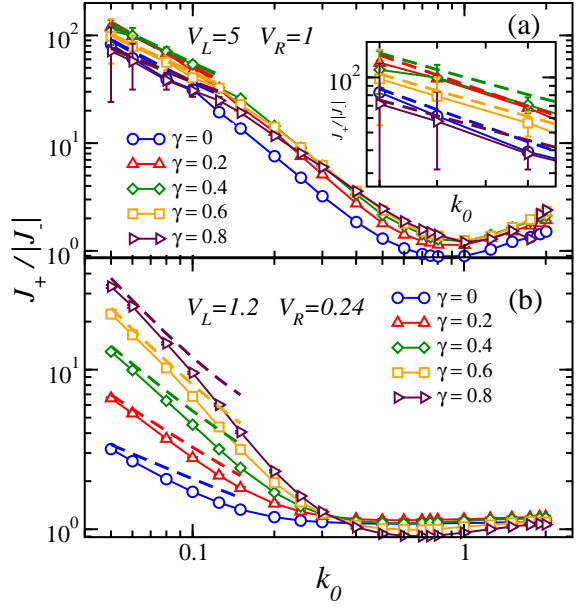}}
\caption{Thermal rectification $r$ vs the interfacial harmonic coupling strength $\ko$ for various values of relative strength of the NNN interactions $\gamma$. In both panels $\lambda=0.2$, but in (a) $\Vl=5$ and in (b) $\Vl=1.2$. All other parameters are the same as those in Fig.~\ref{fig:2}. In both panels dashed lines correspond to the power-law fit $\sim\ko^{-\alpha}$ and in (b) the dashed lines have been risen upwards for clarity. In (a) the inset details the region where the fits were performed and in (b) error bars are smaller than symbol size. Continuous lines are a guide to the eye.}
\label{fig:3}
\end{figure}

The corresponding rectification values obtained for the heat fluxes presented in Fig.~\ref{fig:2} are reported in Fig.~\ref{fig:3}. From the reported results it is clear that, for $\ko<0.1$ in the case of $\Vl=5$ and $\ko<0.15$ for $\Vl=1.2$, the rectification follows a power law behavior $\sim\ko^{-\alpha}$. For the $\Vl=5$ instance reported in panel (a) the exponents are, from low to high $\gamma$ values, $1.5$, $1.4$, $1.3$, $1.2$, and $1.1$ whereas for the $\Vl=1.2$ instance of panel (b) the corresponding values are $0.8$, $1.2$, $1.5$, $1.8$, and $1.9$. The decreasing magnitude of the exponent for $\Vl=5$ as $\gamma$ increases thus signals an increase in TR. On the contrary, for $\Vl=1.2$, $r$ increases as magnitude of the scaling exponent also does so. In this last instance we can infer that, since the difference $\Vl-\Vr=0.96$ is small, the asymmetry afforded by the NNN interactions is the one responsible for the TR effect, and thus $r$ increases as $\lambda$ does so. On the contrary, for $\Vl=5$ ---and thus $\Vl-\Vr=4$--- the asymmetry in the amplitudes of the onsite potential is mostly responsible of the TR effect, and thus the curves in panel (a) have similar behaviors, with scaling exponents closer in magnitude than those of the $\Vl=1.2$ instance.

To study in more detail the phenomenology due to the contribution of the NNN interactions in Fig.~\ref{fig:4} we plot the power spectra $P_i(\omega)=\langle|\tau^{-1}\!\!\int_{_0}^{\tau}\!\! dt\dot q_i(t)\exp(-\mathrm{i}\omega t)|^2\rangle$ of the interface oscillators at the left and right sides of the interface. The Fourier transform is computed over an interval of $\tau=2^{10}$ time units; $\langle\cdots\rangle$ indicates an average over the complete stationary time interval. Considering the results in Fig.~\ref{fig:2}(a), for which $\lambda=0.2$, we choose the value of $\gamma=0.4$ because, for this strength of the NNN interactions, $J_+>J_-$ and thus  the TR value is well defined. First we study the weak-coupling limit with $\ko=0.05$. It is to be noted that, if the temperature on one side of the lattice is above a threshold value $T_{_{\mathrm{cr}}}^{(L,R)}\approx V_{_{L,R}}/(2\pi)^2$, then that side behaves as a harmonic lattice~\cite{Li04a}. For the left side $T_{_L}=0.105\lesssim T_{_{\mathrm{cr}}}^{(L)}=0.13$ for $V_{_L}=5$, and therefore should in principle be in a regime wherein the effect of the onsite potential are relevant. However, it is important to make some important caveats: first, the temperature on the left side is very close to the threshold one, and second it can be readily corroborated in panel (a) that the contribution of low-frequency phonons is small but non-negligible; the latter are associated with a harmonic dynamics. Therefore these facts indicate that the left side is nevertheless in the harmonic regime. Now, when NNN interactions are present the dispersion relation reads as $\omega_{\alpha}=2[(\kl/m)(\sin^2q_{\alpha}/2+\gamma\sin^2q_{\alpha})]^{\frac{1}{2}}$, where $q_{\alpha}$ is the wave number and $\omega_{\alpha}$ the corresponding frequency; the maximum value $\omega_{\mathrm{max}}$ for a certain wave number indicates the phonon frequency limit for the considered side of the system. Therefore, for $\gamma=0.4$ we have $\omega_{\mathrm{max}}/2\pi\sim0.3274$ for $\kl=1$, which define the phonon band $0<\omega/2\pi\lesssim0.3274$ for the left oscillator. For the right side we have $T_{_R}=0.035>T_{_{\mathrm{cr}}}^{(R)}=0.025$ for $V_{_R}=1$ ---thus being clearly in the harmonic regime--- and, due to the absence of NNN interactions, the phonon band is now given by the expression $0<\omega<(4\kr/m)^{\frac{1}{2}}$ which, for $\kr=0.2$, is $0<\omega/2\pi\lesssim0.1466$. Heat conduction can occur only for frequency values in the overlapping region $\Omega/2\pi<0.1466$ of these phonon bands, and there is indeed an appreciable overlap in the aforementioned region when $T_{_L}>T_{_R}$. For the reverse-bias configuration we now have $T_{_{\mathrm{cr}}}^{(L)}=0.13>T_{_L}=0.035$. In this case, being this temperature  well below the threshold value, the influence of the anharmonic FK potential now becomes relevant. In this case the lower bound of the phonon band is raised by $(\Vr)^{\frac{1}{2}}$ and the phonon band is shifted to the maximum value $\omega_{\mathrm{max}}$ obtained from the dispersion relation $\omega_{\alpha}^2=[\Vl + 4\kl(\sin^2q_{\alpha}/2+\gamma\sin^2q_{\alpha})]/m$. For $\Vl=5$ and $\kl=1$ the phonon band is $0.3559<\omega/2\pi<0.4834$, which has no overlap with the right phonon band, as can be appreciated in panel (b). This result explains the high rectification figure observed in Fig.~\ref{fig:3}(a), higher than the corresponding case for $\gamma=0$. For the harmonic coupling strength value $\ko=0.5$ the results for the forward- and reverse-bias configurations are presented in panels (c) and (d), respectively. The phenomenology is similar to the weak-coupling limit already considered, although in this case all the spectra are shifted to higher frequencies. However, this shift has the effect that now there is hardly any spectra overlap in the low-frequency range for the forward-bias configuration, which renders the low rectification value already plotted in Fig.~\ref{fig:3}.

\begin{figure}
\centerline{\includegraphics*[width=85mm]{Fig4.eps}}
\caption{(a) Power spectra of the two oscillators in the left (red) and right (blue) sides of the contact for (a) forward- and (b) reverse-bias configuration with $\Vl=5$, $\gamma=0.4$, and $\ko=0.05$; (c) and (d) the same as (a) and (b) but for $\ko=0.5$. Vertical solid and dot-dashed lines correspond to the cut-off frequencies of the left and right phonon bands respectively. Same $\kl$, $\lambda$, and $N$ values as in Fig.~\ref{fig:2}.}
\label{fig:4}
\end{figure}

\begin{figure}
\centerline{\includegraphics*[width=85mm]{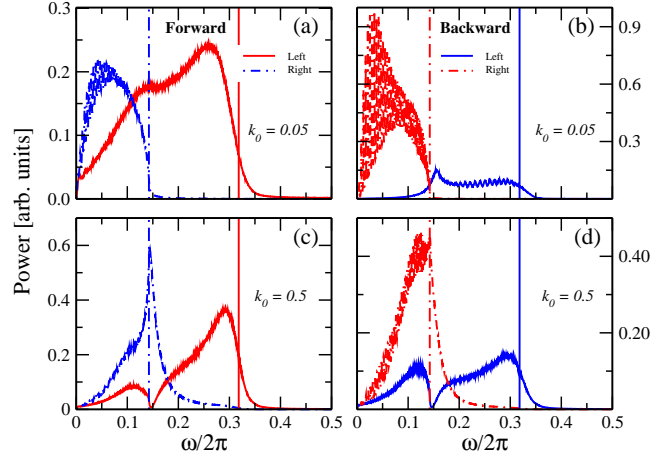}}
\caption{(a) Power spectra of the two oscillators in the left (red) and right (blue) sides of the contact for (a) forward- and (b) reverse-bias configuration with $\Vl=1.2$, $\gamma=0.4$, and $\ko=0.05$; (c) and (d) the same as (a) and (b) but for $\ko=0.5$. Vertical solid and dot-dashed lines correspond to the cut-off frequencies of the left and right phonon bands respectively. Same $\kl$, $\lambda$, and $N$ values as in Fig.~\ref{fig:2}.}
\label{fig:5}
\end{figure}

The results for $\Vl=1.2$ are presented in Fig.~\ref{fig:5}. For this amplitude of the FK potential the critical temperatures for the left and right sides are $T_{_{\mathrm{cr}}}^{(L)}=0.03$ and $T_{_{\mathrm{cr}}}^{(R)}=6\times10^{-3}$ respectively. Therefore in both bias configurations the two halves of the system are in the harmonic regime and the phonon bands correspond to those in panels (a) and (c) of Fig.~\ref{fig:4}. For the $\ko=0.05$ instance depicted in panel (a) corresponding to the forward-bias configuration there is a significant spectral overlap in the low-frequency region, and thus there is heat conduction; for the reverse-bias one significant spectral power is afforded by the right spectrum in the low-frequency region, which increases the heat flux and results in a lower rectification than the corresponding instance for $\Vl=5$. The corresponding cases for the $\ko=0.5$ value presented in panels (c) and (d) clearly show that there is also a large overlap of the spectra in the low-frequency region for both configurations, thus rendering heat fluxes of similar magnitude, as was already evident in Fig.~\ref{fig:2}(a), and a very low rectification figure.

\begin{figure}
\centerline{\includegraphics*[width=85mm]{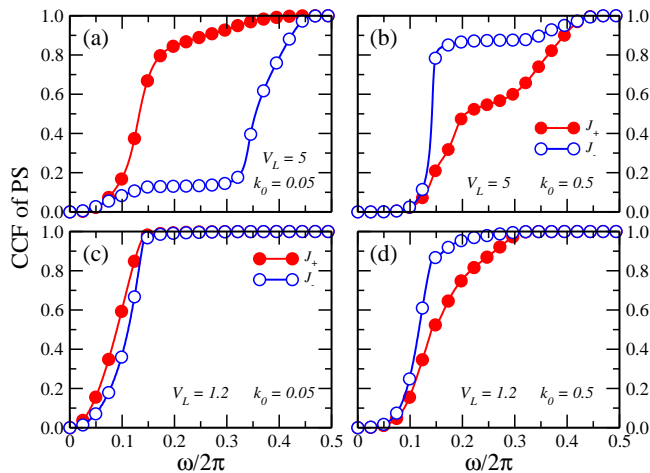}}
\caption{CCF of PS between oscillators $i=50$ and $51$ for (a) $\ko=0.05$ and (b) $\ko=0.5$ with $\Vl=5$; (c) and (d) the same as (a) and (b) but for $\Vl=1.2$. Solid and void symbols correspond to the forward- and reverse-bias configurations, respectively. Same $\kl$, $\lambda$, and $N$ values as in Fig.~\ref{fig:2}. Continuous lines are a guide to the eye.}
\label{fig:6}
\end{figure}

To quantify the degree of overlap of the power spectra between oscillators ---and thus gain further insight into the mechanisms responsible of TR--- the cumulative correlation factor (CCF), introduced in Refs.~\cite{Li05,Zhang17}, is used to represent the match-mismatch degree of vibrational modes among them. The CCF below a specific frequency $\omega_s$ between oscillators $i$ and $j$ is defined as
\be
M_{ij}(\omega_s)={\int_0^{\omega_s}P_i(\omega)P_j(\omega)d\omega\over\int_0^{\infty}P_i(\omega)d\omega\int_0^{\infty}P_j(\omega)d\omega}.
\ee
Each CCF in the two opposite directions is normalized by dividing $M(\omega_s)$ by $M(\infty)$. Previously it has been established that a small CCF at the low frequency range means a larger mismatch, resulting in a lower thermal conductance in that direction~\cite{Dong19}. Thus in Fig.~\ref{fig:6} it can be observed that, for the case $\Vl=5$ and $\ko=0.05$ depicted in panel (a), there is a larger mismatch in the backward direction for low and middle frequency values, which is consistent with the large rectification observed in Fig.~\ref{fig:3}(a). For the $\ko=0.5$ case presented in panel (b), on the contrary, there is a larger mismatch in the forward direction, specially in the intermediate frequency range, which results in a low rectification figure. Next, for $\Vl=1.2$ and $\ko=0.05$ it is clear that the CCFs in the forward and backward directions are very similar, in the latter case due to the large contribution of low-frequency phonons due to the right spectrum, as can be observed in Fig.~\ref{fig:5}(b); therefore, a small rectification is obtained. For $\ko=0.5$ the CCFs are more dissimilar, but the one corresponding to the backward direction has a larger match in the low frequency region that is again produced by the right spectrum that is heavily populated by low-frequency modes, see Fig.~\ref{fig:5}(d), and again a low rectification is obtained.

\subsection{Large-size limit}

\begin{figure}
\centerline{\includegraphics*[width=85mm]{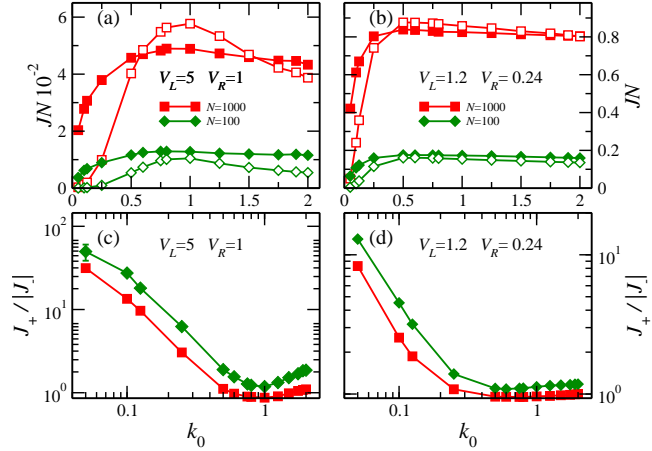}}
\caption{(a,c) Heat flux vs the interfacial harmonic coupling strength $\ko$ for $\gamma=0.4$, $\Vl=5$, and two system sizes, $N=100$ (diamonds) and $1000$ (squares), together with the corresponding rectification; for (b,d) $\Vl=1.2$. Solid and void symbols in (a) and (b) correspond to the forward- and reverse-bias configurations, respectively. $\kl=1$ and $\lambda=0.2$ in all instances. Error bars are smaller than system size. Continuous lines are a guide to the eye.}
\label{fig:7}
\end{figure}

In Fig.~\ref{fig:7}(a) we present the results for the total heat flux $JN$ in the forward- and reverse-bias configurations as a function of the harmonic coupling strength $\ko$ for $\gamma=0.4$ and $\Vl=5$ previously shown in Fig.~\ref{fig:2}(a) alongside the same result for $N=1000$. It can be observed that the reversal of the TR effect, i.e $J_->J_+$, now appears for this large system size. However it is important to note that, for this very same system size of $N=1000$ and without NNN interactions, this effect was way bigger and appeared for small $\ko\sim0.25$ values~\cite{Hu06}, whereas in our case appears at larger $\ko\sim0.5$ values and is much smaller in magnitude. The corresponding TR, depicted in Fig.~\ref{fig:7}(c), presents the expected decay in rectification efficiency for $N=1000$ in the range $\ko<0.5$. For $\Vl=1.2$ the heat fluxes for both system sizes, presented in Fig.~\ref{fig:7}(b), have a weak dependence on the harmonic coupling strength for $\ko>0.5$. In the $N=1000$ instance the TR reversal is smaller than the reversal already noted for $\Vl=5$. Therefore it can be stated that a strong asymmetry is needed to maintain a sizable difference of the heat fluxes in both configurations for small system sizes. For larger $N$ values the TR reversal appears, but its magnitude is greatly diminished in comparison to the case with NN interactions only and anyway appears in a $\ko$ range wherein TR becomes insignificant in magnitude.

\begin{figure}
\centerline{\includegraphics*[width=85mm]{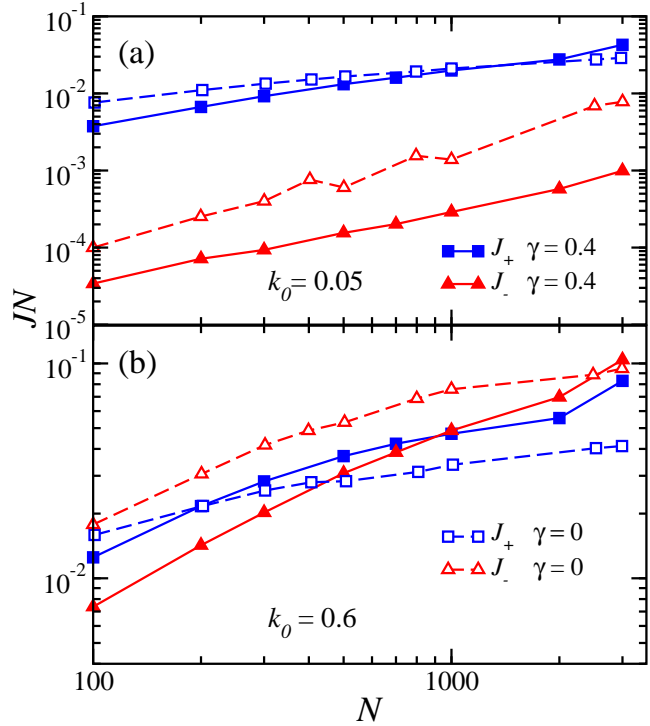}}
\caption{Dependence of the total heat flux $J_{_\pm}N$ on the system size $N$ for (a) $\ko=0.05$ and (b) $\ko=0.6$ with $\Vl=5$, $\kl=1$, $\lambda=0.2$, and $\gamma=0.4$. Void symbols correspond to $\gamma=0$ in both panels. Continuous and dashed lines are guides to the eye. Error bars are smaller than the symbol size.}
\label{fig:8}
\end{figure}

Fig.~\ref{fig:8} presents the dependence of the total heat flux $JN$ on the system size $N$ for $\ko$ values in the weak-coupling limit and above, again for $\gamma=0.4$ and $\Vl=5$. For the case without NNN interactions it was determined that a crossover wherein $J_->J_+$ occurs at $N\sim3000$, with $J_-$ significantly increasing as $N$ grows~\cite{Li04a}. However in Fig.~\ref{fig:8}(a) the phenomenology for $\gamma=0.4$ is rather different: the difference $\Delta J\equiv J_+-|J_-|$ has a weak dependence on the system size for all the considered $N$ values, which stands in sharp contrast to the corresponding case $\gamma=0$ displayed in that same figure that indicates that $\Delta J$ diminishes as $N$ increases. Thus, although a crossover for even larger $N$ values cannot be completely ruled out by our results, it is clear that $J_+>J_-$ will remain valid for $N\gg3000$, contrary to the case without NNN interactions. For a $\ko$ value beyond the weak-coupling limit the results presented in Fig.~\ref{fig:8}(b) indicate that the crossover occurs at $N\sim1000$, in sharp contrast with the corresponding results without NNN interactions that clearly indicate that $J_->J_+$ for all $N$ values considered. Therefore, the presence of NNN interactions improves the performance of the device as a rectifier for a larger range of $N$ values than the corresponding case with NN interactions only in the weak-coupling limit. Beyond the aforementioned limit the heat flux values are very similar for $N<1000$ and thus the system cannot function as a rectifier.

\subsection{Local heat flux}

\begin{figure}
\centerline{\includegraphics*[width=85mm]{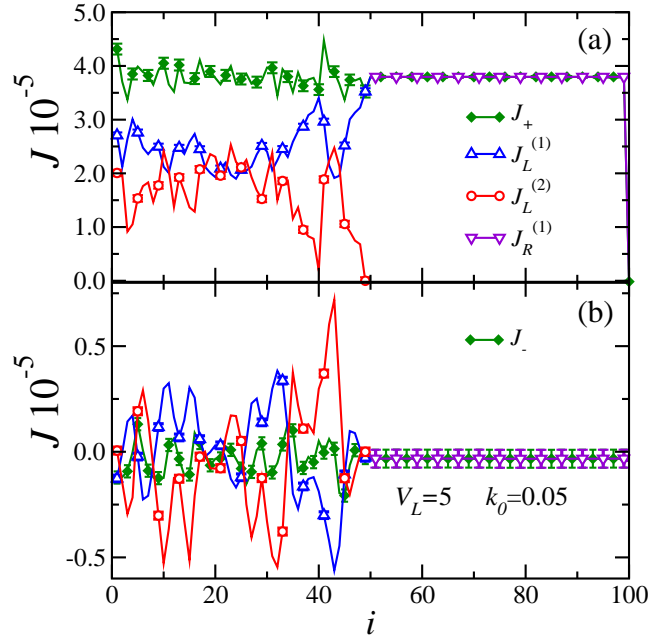}}
\caption{Local heat flux for a lattice with $N=100$, $\kl=1$, $\Vl=5$, $\lambda=0.2$, $\gamma=0.4$, and $\ko=0.05$; Dashed lines with triangles indicate the NN contribution, dot dashed lines with circles the NNN one, and solid lines the total heat flux for (a) $T_{_L}>T_{_R}$ and (b) $T_{_L}<T_{_R}$.}
\label{fig:9}
\end{figure}

In Fig.~\ref{fig:9} we plot separately the contribution of the NN term $J_i^{(1)}$ to the local heat flux, that of the NNN one $J_i^{(2)}$, and the total local heat flux $J_i$ given by Eq.~\ref{lhf} for a lattice with $\Vl=5$ and $\gamma=0.4$ in the weak-coupling limit. For the forward-bias configuration depicted in panel (a) it can be readily observed that, although on average the local heat flux is constant along the entire length of the system, as expected, on each side its behavior is rather different. On the right side the variation of the local heat flux from site to site is negligible, whereas in the left one where the NNN interactions are present the variations in the local heat flux are of significant magnitude. These are not a finite-time effect since the reported results in this and the next figure were obtained employing a stationary time interval of $2\times10^8$ time units, one order of magnitude greater than that employed to obtain the rest of our results; the differences with those obtained with $2\times10^7$ (not shown) are negligible. The spatial averages of both local heat fluxes are $J_+^{(1)}=2.3\times10^{-5}$ and $J_+^{(2)}=1.8\times10^{-5}$, with a total magnitude for the local heat flux of $J_{+}=3.7\times10^{-5}$. The case of the reverse-bias configuration has some significant differences. The local fluxes corresponding to the NN and NNN interactions have larger variations than the corresponding ones of the $\Tmpl>\Tmpr$ configuration. On average the NNN contributions seem to be the ones that drive the total heat flux to a negative average value, since the NN ones have a positive contribution in more sites than the NNN ones. These observations are corroborated by the average spatial values of each heat flux, which are $J_i^{(1)}=1.1\times10^{-6}$, $J_i^{(2)}=-1.8\times10^{-6}$, with a local heat flux of $J_{-}=-3.4\times10^{-7}$. Therefore the effect of the large onsite potential amplitude in the left side is to render a large rectification, but with a rather non-uniform behavior of the local heat flux in that same side.

\begin{figure}
\centerline{\includegraphics*[width=85mm]{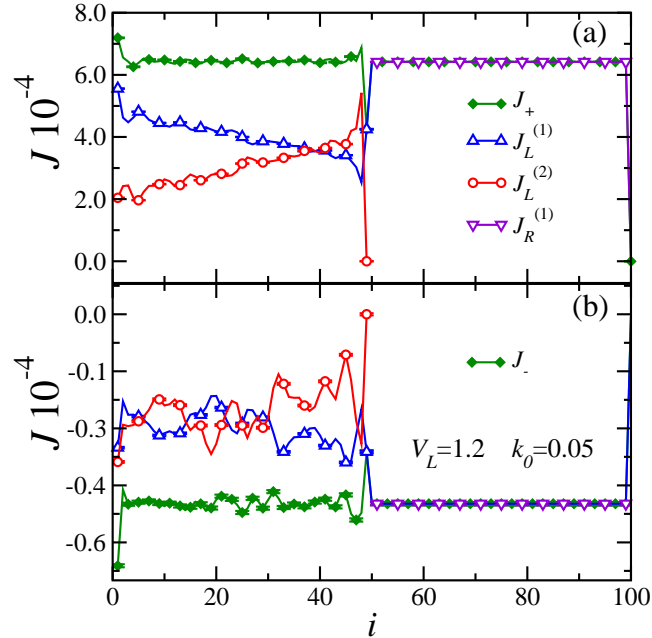}}
\caption{Same as Fig.~\ref{fig:7}, but now for a lattice with $\Vl=1.2$. (a) $T_{_L}>T_{_R}$ and (b) $T_{_L}<T_{_R}$. Error bars are smaller than system size.}
\label{fig:10}
\end{figure}

In Fig.~\ref{fig:10} we present the same results as in Fig.~\ref{fig:9} but now for a system with $\Vl=1.2$. For $T_{_L}>T_{_R}$ the local contribution to the heat flux of the NN term decreases in the direction of the the total heat flow along the lattice, whereas the contribution arising form the NNN interactions increases, both in a linear way. In this case, if we again take the spatial average, we obtain $J_+^{(1)}=4.0\times10^{-4}$, $J_+^{(2)}=2.9\times10^{-4}$, with $J_+=6.3\times10^{-4}$. We notice that the smooth behavior of $J_i^{(1)}$ and $J_i^{(2)}$ is identical to the one observed in the corresponding heat fluxes along the entire length of a mass-graded anharmonic lattice with NN and NNN interactions but without an onsite potential~\cite{Romero17}. When the thermal reservoirs are swapped the same behavior is observed on average, but now there is more variation in the local heat flux from site to site. The corresponding spatial averages are $J_-^{(1)}=-2.9\times10^{-5}$ and $J_-^{(2)}=-2.4\times10^{-5}$, with a total heat flux of $J_-=4.9\times10^{-5}$. For the employed value of $\Vl=1.2$ both sides of the system are more similar to each other and the asymmetry afforded by the NNN interactions becomes more relevant. Furthermore, since the system is in the weak-coupling limit, the dynamical behavior of both sides is largely independent from one another. Therefore, when the hot reservoir is directly in contact with the side where NNN interactions exist the influence of the right side becomes subsumed with that of the cold reservoir. Thus the behavior of $J_+^{(1)}$ and $J_+^{(2)}$ on the left side is very similar to the one observed in the system without an interface previously mentioned. On the other hand, for the reverse-bias configuration the side with the NNN interactions is now in contact with the cold reservoir, and thus the hot reservoir only interacts indirectly with it. In this configuration the hot reservoir cannot smooth the behavior of the local heat flux on the left side as it seems to be able to do so in the forward-bias configuration. The final result is then a non-smooth behavior in the local heat flux.

\subsection{Dependence on the asymmetry parameter}

\begin{figure}\centering
\includegraphics[width=0.65\linewidth,angle=0.0]{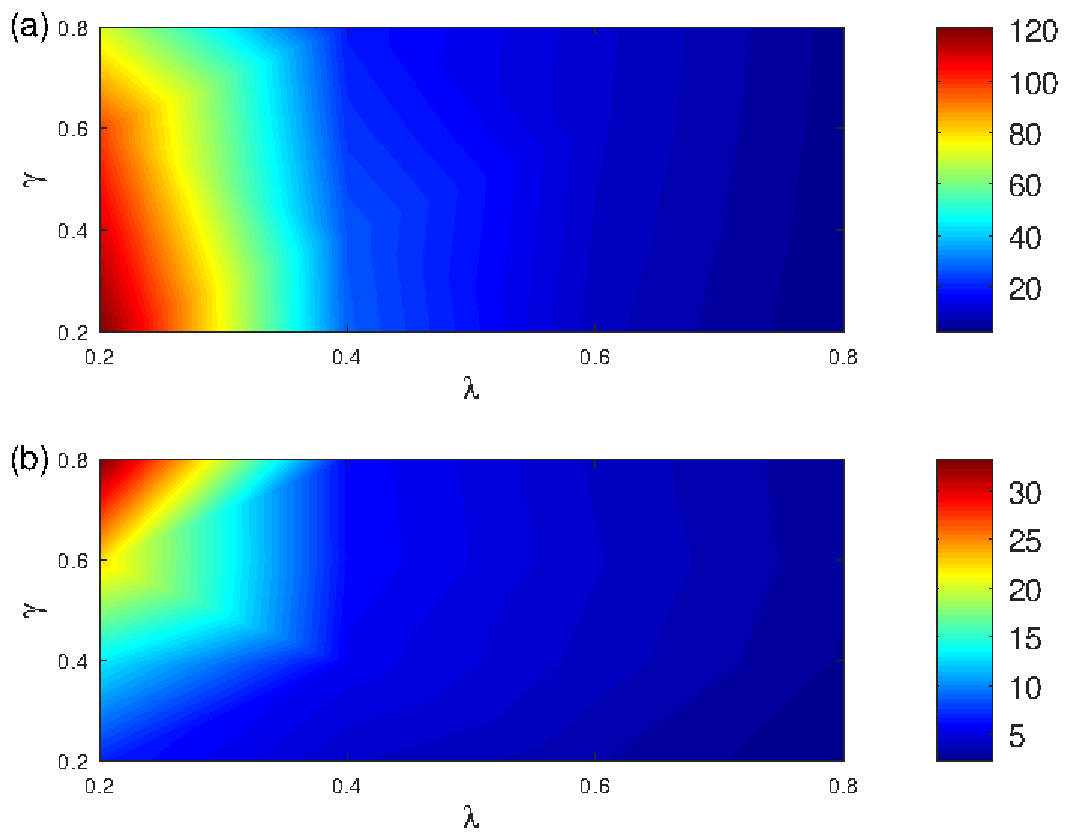}
\caption{Surface plot of the thermal rectification $r$ as a function of the asymmetry parameter $\lambda$ and the relative strength of the NNN interactions $\gamma$. For (a) $\Vl=5$, $\kl=1$, $\ko=0.05$, and $N=100$; for (b) $\Vl=1.2$, $\kl=1$, $\ko=0.05$, and $N=100$ .}
\label{fig:11}
\end{figure}

Next we will explore the properties of the system as a function of the scaling parameter $\lambda$, which controls the effects of changes in both the harmonic constant and the strength of the onsite potential; for the latter we consider first the $\Vl=5$ value. In Fig.~\ref{fig:11} (a) we report the surface plot of the TR as a function of the asymmetry parameter $\lambda$ and the relative strength of the NNN interactions $\gamma$. It can be readily appreciated that the highest rectification values are obtained within the region defined by $\lambda<0.4$ and $\gamma<0.6$; for larger $\lambda$ values the aforementioned TR effect is greatly diminished. For low $\lambda$ values TR diminishes as the relative strength of the NNN interactions increases. Now, for the $\Vl=1.2$ case presented in Fig.~\ref{fig:11} (b) it can be noticed that the region of high TR is obtained for very high asymmetry, i.e. low $\lambda$ values, and very high values of the relative strength of the NNN interactions $\gamma$. This last result can be accounted for by the fact that, in this case, the difference $\Vl-\Vr=0.96$ is small and thus the asymmetry on the type of interactions at each side of the system is the main structural asymmetry to account for the moderate values of the obtained TR.

\begin{figure}\centering
\includegraphics[width=0.65\linewidth,angle=0.0]{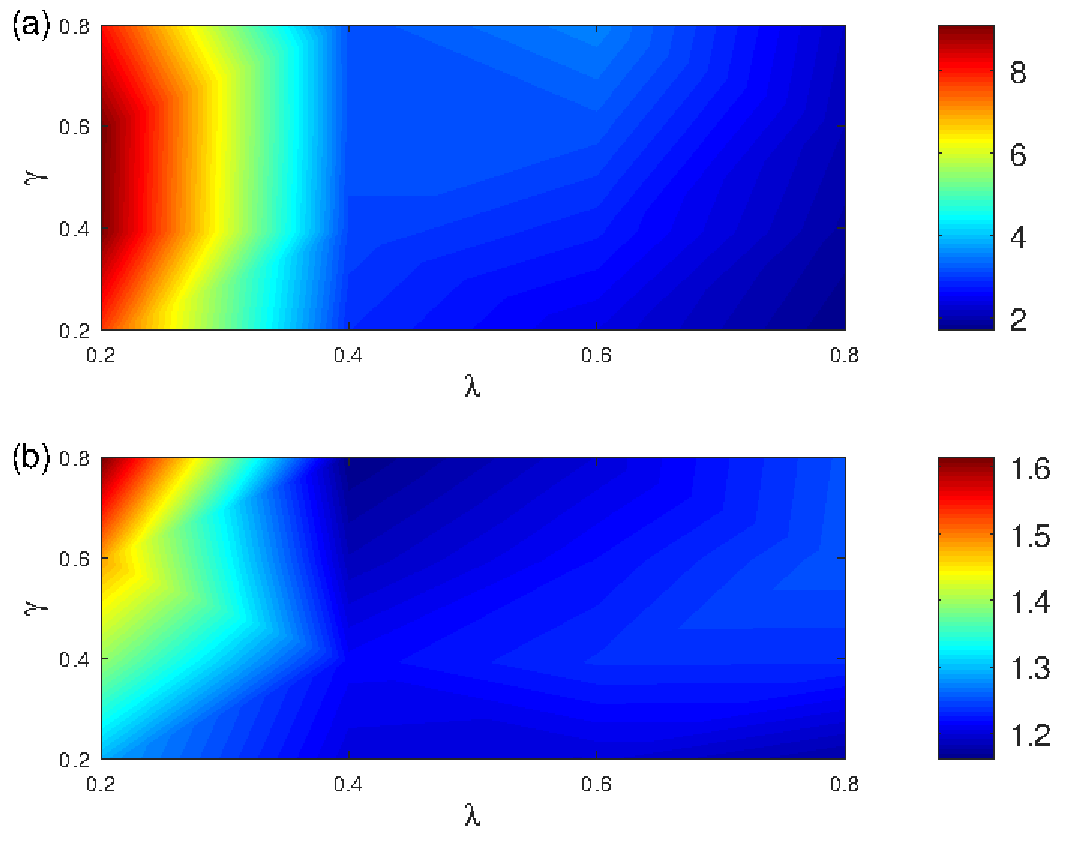}
\caption{Same as Fig.~\ref{fig:11}, but now for an interfacial harmonic coupling constant of $\ko=0.25$.}
\label{fig:12}
\end{figure}

Finally in Fig.~\ref{fig:12} we present the surface plots of the TR for both amplitude values of the left onsite potential, but now for $\ko=0.25$. The phenomenology is very similar to that corresponding to $\ko=0.05$ already presented in Fig.~\ref{fig:11}, but with much lower rectification values. A possible explanation, obtained for this rather large value of the interfacial harmonic coupling strength, is that the surface plots (not shown) for both forward- and reverse-bias configurations are very similar to each other for all combinations of $\lambda$ and $\gamma$ values, precluding any sizable rectification figure.

\section{Concluding remarks\label{sec:Disc}}

We have investigated the influence of NNN interactions on the TR phenomenon present in a 1D system composed of two dissimilar FK lattices connected by a weak harmonic coupling. The NNN interactions afford a first step to consider the effects on TR of the long-range interactions, such as of the Morse and Lennard-Jones type, present in molecular junctions consisting, for example, of polyethylene polymer chains placed between and attached to metal substrates subject to a static thermal bias~\cite{Dinpajooh22}. Our results confirm that TR is stronger in the weak-coupling limit, just as in the previously considered case of NN interactions alone. However, the reversal of the TR effect ---which greatly diminished the efficiency of the device as a rectifier when only NN interactions are present--- is strongly diminished when NNN interactions of moderate magnitude exist on the left side of the system. For the case of a more symmetric system TR obviously diminishes compared to the more asymmetric case. However, for this latter instance TR is significantly improved, albeit within a diminished TR efficiency range, by the presence of NNN interactions. Furthermore, within the weak-coupling limit, TR has a sizable value ---$r\sim43$ for $N=3000$--- as the system size increases in the presence of NNN interactions. Thus it seems that these type of interactions help in an important way to stabilize the behavior of the system under structural changes ---variations in the $\ko$ value--- that seemed to render the proposed system unsuitable as a thermal rectification device when only NN interactions were considered~\cite{Hu06}. The spectral analysis reveals that, for the instance with the highest asymmetry in the amplitudes of the onsite potential, the behavior of the power spectra is remarkably similar both for low and high values of the harmonic coupling constant, which certainly helps to explain why TR is larger for the NNN instance compared to the one with only NN interactions when the system is beyond the weak-coupling limit. 

Finally we would like to emphasize that our results may be relevant to explain those recently reported for MBs. For example, a significant TR effect was measured in MBs that covalently bond a gold lead and a carbon nanotube, where the structural asymmetry is afforded by the large mass gradients formed between the metal leads and the MBs or inside the elements of the MBs~\cite{Dong19}. However, in a later study of MBs composed of an alkane junction and metal leads the TR effect was inconsequential ---unless one goes to very large thermal biases~\cite{Wang22}---. This finding suggests that one should seek molecules with considerable internal anharmonic effects for developing thermal devices. In our system the anharmonicity is controlled by means of the interaction with a substrate. Therefore this mechanism offers a possible means of implementing the anharmonicity needed to obtain TR in this type of systems.

\ack
M.~R.~B. thanks Consejo Nacional de Humanidades, Ciencias y Tecnolog\'\i as (CONAHCYT) Mexico for financial support. A.~P.~V. thanks ``Programa Institucional de Formaci\'on de Investigadores" I.P.N. M\'exico for financial support.


\section*{References}
\bibliography{bibliografia}

\end{document}